\documentclass[%
 reprint,
superscriptaddress,
bibnotes,
 amsmath,amssymb,
 aps,
]{revtex4-1}

\usepackage{graphicx}
\usepackage{dcolumn}
\usepackage{bm}
\usepackage[T1]{fontenc}
\usepackage{mathptmx}
\usepackage{mathrsfs}
\usepackage{xcolor}
\usepackage{hyperref}
\usepackage{physics}
\usepackage{simplewick}
\usepackage{mathrsfs}
\usepackage[bottom]{footmisc}
\usepackage{threeparttable}

\newcommand{\be}{\begin{equation}}
\newcommand{\ee}{\end{equation}}
\newcommand{\bs}{\begin{split}}
\newcommand{\es}{\end{split}}

\begin{document}

\title{Application of optomechanical frequency conversion on gravitational wave detection}
\author{Yiqiu Ma}
\email{myqphy@gmail.com}
\affiliation{Center for gravitational experiment, School of Physics, Huazhong University of Science and Technology, Wuhan, 430074, Hubei, China}
\affiliation{Theoretical Astrophysics 350-17, California Institute of Technology, Pasadena, CA 91125, USA}
\author{Feng-Li Lin}
\affiliation{Department of Physics, National Taiwan Normal University No.88, Sec. 4, Ting-Chou Road, Taipei, 11677, Taiwan}
\author{Haixing Miao}
\affiliation{School of Physics and Astronomy, Institute of Gravitational Wave Astronomy, University of Birmingham, Birmingham B15 2TT, United Kingdom}
\author{Chunnong Zhao}
\affiliation{The University of Western Australia, School of Physics (OzGrav-UWA), WA6011, Australia}
\author{Yanbei Chen}
\affiliation{Theoretical Astrophysics 350-17, California Institute of Technology, Pasadena, CA 91125, USA}

\begin{abstract}
Optomechanical interaction can be a platform for converting quantum optical sate between different frequencies.
In this work, we propose to combine the idea of optomechanical frequency conversion and the dual use of laser interferometer, for the purpose of improving the sensitivity of laser interferometer gravitational wave detectors by filtering the light field. We found that compare to the previous schemes of implementing the optomechanical devices in gravitational wave detectors, this frequency converter scheme will have less stringent requirement on the thermal noise dilution.
\end{abstract}
\maketitle

\section{Introduction}
LIGO's first detection of gravitational waves emitted from the binary black hole system GW150914 opens the era of gravitational wave (GW) astronomy\,\cite{GW150914}. The dominate noise source of the advanced interferometric GW detectors, e. g. the advanced LIGO, advanced VIRGO and KAGRA is quantum noise, over almost the entire detection band\,\cite{Adhikari2014, Martynov2017}. The origin of quantum noise is the vacuum fluctuation of electromagnetic fields\,\cite{Kimble2001}. Concretely, the vacuum phase fluctuations (shot noise) and amplitude fluctuations (radiation pressure noise) dominate at relatively high frequencies and low frequencies, respectively. The trade-off between these two contributions is the Standard Quantum Limit (SQL)\,\cite{Kimble2001}.

Further improvement of the detector sensitivity requires the surpassing of SQL by engineering the quantum state of light. Two typical ways of surpassing SQL are (1) frequency-dependent (FD) squeezed light injection\,\cite{Kimble2001} and (2) FD (variational) readout\,\cite{Kimble2001}. The first method makes use of a filtered squeeze light with reduced phase/amplitude uncertainty at the high/low frequencies, respectively; while the second method filters the output light in a FD way so that the radiation pressure noise diminishes at the homodyne detector. Both methods require a narrowband filter cavity (e.g. $\sim 50$Hz for AdvLIGO). This means either the length of the filter cavity is long ($\sim 10^2-10^3$ meters) or the optical loss of the filter cavity must be small.  Currently for FD squeezing injection, KAGRA proposes to build 300 meter filter cavities\,\cite{Caposcasa2016} and the work of testing for 16 meter filter cavity is on-going\,\cite{Oelker2016,Evans2013,Isogai2013}.

Various other ways are also proposed to achieve broadband squeezing. There are two main approaches: (1) using the interferometer itself as a filter cavity\,\cite{Ma2017,Barsotti2014}; (2) replacing the long cavity by some table-top optomechanical devices with a similar dispersion behavior\,\cite{Ma2014}.  Moreover, table-top optomechanical devices could find their applications not only for beating SQL, but also for enhancing the detection bandwidth\,\cite{Miao2015PRL}. Notably, recent remarkable experiment of observing quantum radiation pressure effect  at room temperature for an table-top optomechanical device in\,\cite{Cripe2018} shed light on the prospect of implementing optomechanical filters in future GW detection.

However, it is important to note that optomechanical devices in these designs meet the stringent challenge from thermal noise, estimated as\,\cite{Ma2014}
\be\label{eq:criterion}
\frac{T}{Q_m} \ll \frac{\hbar\gamma_{\rm opt}}{k_B},
\ee
where $T$ and $Q_m$ is the environmental temperature and the quality factor of the mechanical oscillator in the optomechanical device, respectively. \emph{The $\gamma_{\rm opt}$ measures the strength of optomechanical interaction and relates to some typical bandwidth of the main interferometer}. For example, for creating a FD filtering of squeezed light, $\gamma_{\rm opt}\sim 50{\rm Hz}$. Physically this condition Eq.\,\eqref{eq:criterion} means that the thermal force noise acting on the mechanical oscillator must be smaller than the quantum radiation pressure force noise.  A simple estimation can show that in these designs, $T/Q_m\sim 10^{-10}$\,K\,\cite{Ma2014}.

In this work, we propose an alternative method to realise optical filtering, combining the above two approaches. The basic idea is that the main interferometer is proposed to be dual-used, to be a GW detector at one frequency and an optical filter at a far-detuned frequency. The conversion between these two different frequencies are realised by an optomechanical frequency converter (OMFC)\,\cite{Tian2010,Hill2012,Lecocq2016}. As we will see later, the above Eq.\,\eqref{eq:criterion} can be relieved in this design since in this case \emph{the $\gamma_{\rm opt}$ is a quantity independent from main interferometer parameters}.

In Section II, we first give a brief discussion of the physical principles of OMFC. Then in Section III, two configurations of GW detector with OMFC for doing FD squeezing injection and variational readout are discussed. Section IV devotes to the analysis of the effect of various noise and imperfections to the sensitivity. We then conclude with a brief summary of our result and some discussion on the future experiment.

\section{Optomechanical frequency conversion}
The frequency conversion between two optical degrees of freedom (d.o.f) mediated by the mechanical d.o.f is schematically shown in Fig.\,\ref{fig:scheme}. Two optical cavities with resonant frequencies $\omega_a$ and $\omega_c$ are connected by an oscillating reflective mirror with resonant frequency $\omega_m$ and quality factor $Q$. These two cavities are pumped by classical coherent light $\bar a$ and $\bar c$ red-detuned with respect to $\omega_a$ and $\omega_c$ by $\omega_m$, respectively. The probe field (say a squeezed light) with center frequency $\omega_a$ carrying the quantum information at its sidebands enters the left cavity, beats with coherent light $\bar a$. This beating creates a radiation pressure force oscillating $\sim \omega_m$, which effectively drives the motion of the mechanical oscillator.  This is the so-called ``writing process" during which quantum information is written on the mechanical oscillator. Then the motion of mechanical oscillator will modulate the coherent field $\bar c$ in the right cavity, create anti-Stokes and Stokes sidebands with frequency near $\omega_c$ and $\omega_c-2\omega_m$, respectively.  The Stokes sideband is far-off resonance and thereby suppressed.  Finally the outgoing field is a squeezed light with centre frequency at $\omega_c$. This is the so-called ``reading out" process during which the quantum information stored on the mechanical oscillator will be carried out by the field centred at $\omega_c$, thereby completed the frequency conversion process.

\begin{figure}[htbp]\label{fig:scheme}
   \centering
   \includegraphics[width=3.5in]{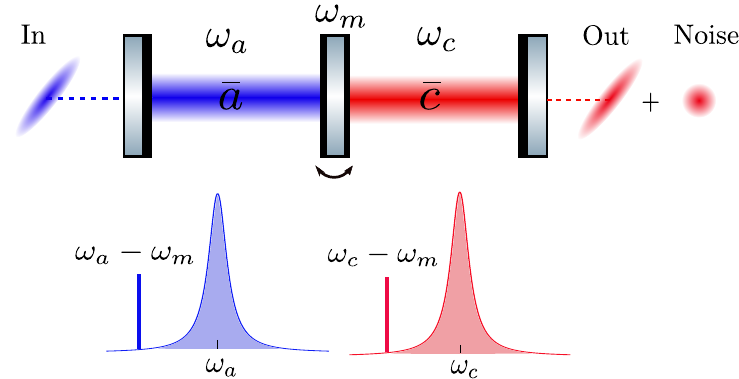} 
   \caption{Schematics of the coupled cavity setup for the optomechanical frequency converter (OMFC). These cavities are coupled individually by two red-detuned coherent light with detuning equal to the mechanical frequency.}
   \label{fig:FC_scheme}
\end{figure}

We can establish a simple Hamiltonian to describe this system:
\be
\begin{split}
&\hat H=\hat H_0+\hat H_{\rm int}+\hat H_{\rm ext},\\
&\hat H_0=\hbar \omega_a\hat a^\dag\hat a+\hbar \omega_c\hat c^\dag\hat c+\hbar \omega_m\hat b^\dag\hat b,\\
&\hat H_{\rm int}=\hbar (G_a\hat a^\dag\hat a-G_c\hat c^\dag \hat c)\hat x,\\
&\hat H_{\rm ext}=i\hbar(\sqrt{2\gamma_a}\hat a^\dag\hat a_{\rm in}+\sqrt{2\gamma_c}\hat c^\dag\hat c_{\rm in}-{\rm h.c}).
\end{split}
\ee
Here, the $\hat a$, $\hat c$, $\hat b$ are annihilation operator of the cavity field in left, right cavities and the mechanical quanta, respectively. The $\hat H_{\rm int}$ describes the radiation pressure forces exert on the mechanical oscillator, note that the minus sign is due to the force contributed by the left and right cavity mode has opposite directions. The $G_{a,c}=\omega_{a,c}/L_{a,c}$ is the single-photon optomechanical coupling constant, where $L_{a,c}$ are the length of left and right cavity. The $\hat H_{\rm ext}$ describes the interaction between the cavity mode and external continuum.

The above interaction can be further simplified. First, the field operators above can be written in the rotation frames as: $\hat a\rightarrow\hat a e^{-i\omega_a t}$, $\hat c\rightarrow \hat c e^{-i\omega_c t}$, $\hat b\rightarrow \hat b e^{-i\omega_m t}$ and thereby $\hat x(t)=x_{\rm zpf}(\hat b e^{-i\omega_m t}+\hat b^\dag e^{i\omega_m t})$ where $x_{\rm zpf}=\sqrt{\hbar/(2m\omega_m)}$ is the ground state displacement of the mechanical oscillator. Since the pumping field is a strong coherent beam, the interaction Hamiltonian for the left cavity $\hat a$ can also be written as:
\be
\begin{split}
\frac{\hat H_{a\rm int}}{\hbar G_0}=&(\bar a e^{-i(\omega_a-\omega_m)t}+\delta\hat a^\dag e^{i\omega_at})(\bar a e^{i(\omega_a-\omega_m)t}+\delta \hat a e^{-i\omega_at})\\
&\times x_{\rm zpf}(\hat b e^{-i\omega_m t}+\hat b^\dag e^{i\omega_m t}),
\end{split}
\ee
where $\bar a$ is the classical pumping field amplitude of left cavity mode, and the $\delta \hat a\ll \bar a$ describes its perturbation due to mechanical modulation or coupling with external continuum. In the future, we will write $\delta \hat a$ as $\hat a$ for convenience. Clearly,  a similar simplification can also be done for $\hat c$. 

Neglecting the non-rotating wave part and the high order perturbation term, the Hamiltonian can be simplified as:
\be
\hat H_{\rm int}=\hbar \bar G_a \hat a\hat b^\dag-\hbar \bar G_c \hat c\hat b^\dag+{\rm h.c}.
\ee
where $\bar G_{a}=G_{a}\bar a x_{\rm zpf}$ ($G_c$ is defined in the same way).  The equations of motion of this simplified Hamiltonian are:
\be
\begin{split}
&\dot{\hat a}=-\gamma_a\hat a-i\bar G_a\hat b+\sqrt{2\gamma_a}\hat a_{\rm in},\\
&\dot{\hat c}=-\gamma_c\hat c+i\bar G_c\hat b+\sqrt{2\gamma_c}\hat c_{\rm in},\\
&\dot{\hat b}=-\gamma_m \hat b -i \bar G_a\hat a+i\bar G_c\hat c+\sqrt{2\gamma_m}\hat b_{\rm th}.
\end{split}
\ee
where $\hat b_{\rm th}$ describes the thermal bath.

As a first step, we consider the ideal case that we ignore the $\gamma_m$ and its associated thermal noise. Besides, we assume that the time scales for the $\hat a,\hat c$ dynamics are much longer than $1/\gamma_{a,c}$ so that $\hat a,\hat c$ can be adiabatically eliminated, then the above equation for $\hat b$ can be reduced to:
\be
\dot{\hat b}=-(\gamma_{\rm opt a}+\gamma_{\rm optc})\hat b-i\sqrt{2\gamma_{\rm opta}}\hat a_{\rm in}+i\sqrt{2\gamma_{\rm opt c}}\hat c_{\rm in},
\ee
where $\gamma_{\rm opt a,c}=\bar G_{a,c}^2/\gamma_{a,c}$ which is the cooling rate of the mechanical oscillator. With this equation of motion for $\hat b$, the output field $\hat a_{\rm out}, \hat c_{\rm out}$ can be solved in the frequency domain as:

\be\label{eq:conversion}
\begin{split}
\hat c_{\rm out}=\frac{\gamma_{\rm opta}-\gamma_{\rm optc}-i\Omega}{\gamma_{\rm opta}+\gamma_{\rm optc}-i\Omega}\hat c_{\rm in}+\frac{2\sqrt{\gamma_{\rm opta}\gamma_{\rm optc}}}{\gamma_{\rm opta}+\gamma_{\rm optc}-i\Omega}\hat a_{\rm in},\\
\hat a_{\rm out}=\frac{\gamma_{\rm optc}-\gamma_{\rm opta}-i\Omega}{\gamma_{\rm opta}+\gamma_{\rm optc}-i\Omega}\hat a_{\rm in}+\frac{2\sqrt{\gamma_{\rm opta}\gamma_{\rm optc}}}{\gamma_{\rm opta}+\gamma_{\rm optc}-i\Omega}\hat c_{\rm in}.
\end{split}
\ee\\

The parameters can be chosen in a way that $\gamma_{\rm opta}=\gamma_{\rm optc}=\gamma_{\rm opt}\gg\Omega$, and the above in-out relation can be further simplified as:
\be\label{eq:conversion_approx}
\left(
\begin{array}{c}
\hat c_{\rm out}\\
\hat a_{\rm out}
\end{array}
\right)=
\left(
\begin{array}{cc}
-i\Omega/2\gamma_{\rm opt}&1\\
1&-i\Omega/2\gamma_{\rm opt}
\end{array}
\right)
\left(
\begin{array}{c}
\hat c_{\rm in}\\
\hat a_{\rm in}
\end{array}
\right),
\ee
which clearly reflects the frequency conversion effect: in the ideal case, any squeezed field with sidebands distributed around $\omega_a$ will be converted to the same squeezed field with sidebands distributed around $\omega_c$.

\begin{figure}[htbp]
   \centering
   \includegraphics[width=2.7 in]{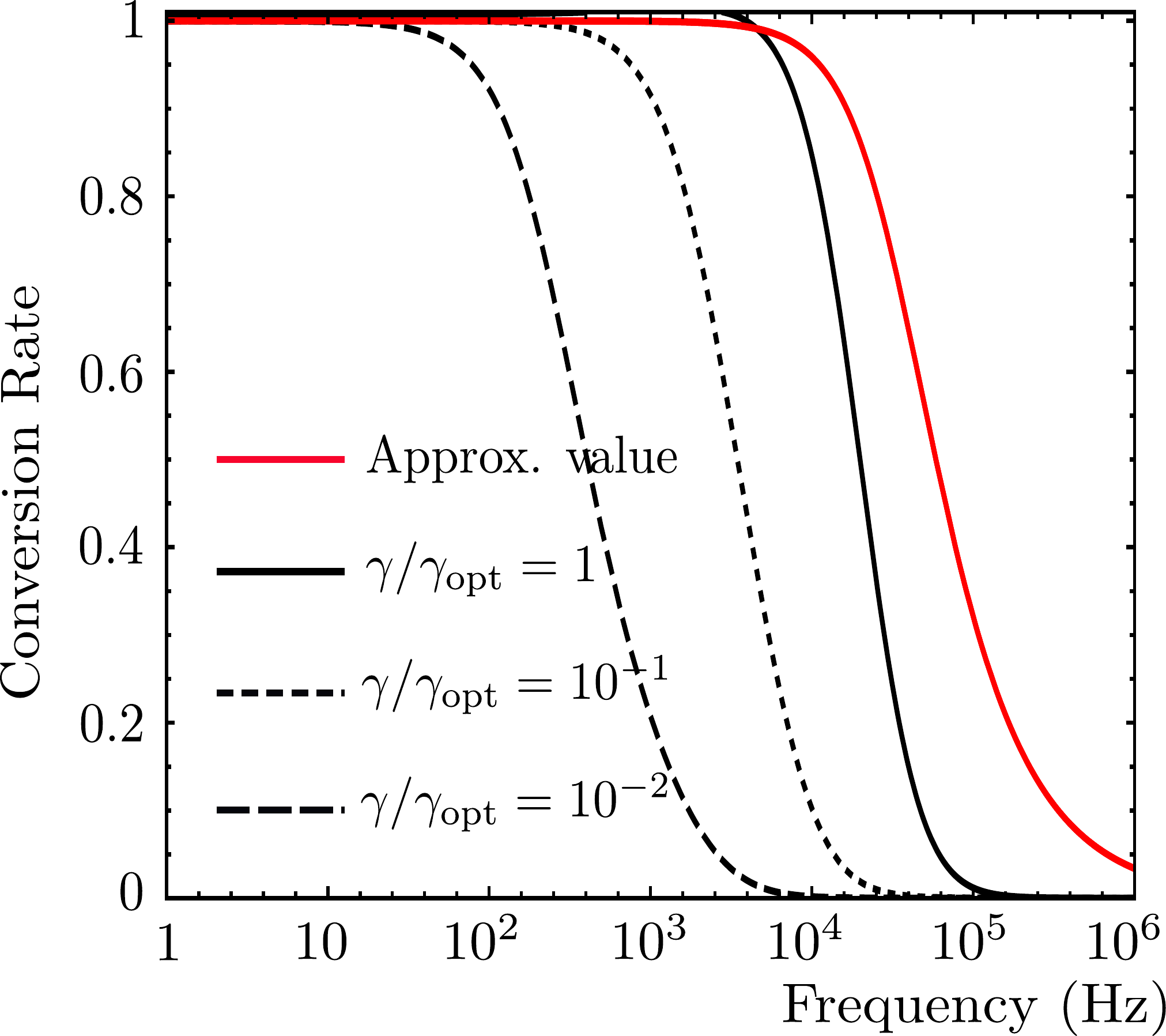} 
   \caption{Conversion rate of OMFC vs frequency. At low frequency region, the conversion rate is almost ideally equal to one since $\Omega\ll\gamma,\gamma_{\rm opt}$. When the ratio $\gamma/\gamma_{\rm opt}$ decrease, this ideal conversion rate will be limited by the cavity bandwidth.}
   \label{fig:omfc}
\end{figure}

Exact and approximated value of conversion coefficients defined as the non-diagonal term in \,\eqref{eq:conversion_approx} are shown and compared in Fig.\,\ref{fig:omfc}, using the parameters shown in Tab,\,\ref{tab:parameters}, from which we can see that the conversion is constrained by the cavity bandwidth. Intuitively, the conversion happens due to the coupling of mechanical oscillator with the intra-cavity fields, which is shaped by the cavity profile. This also explains the squeezing degradation at high frequency in Fig.\,\ref{fig:fd_sensitivity} of the next section.

Among the various noises that affect this frequency conversion,  the most important one is the thermal noise given by:
\be\label{eq:conversion_thermal}
\begin{split}
&\hat c_{\rm out}^{\rm th}(\Omega)=\frac{2i\sqrt{\gamma_m\gamma_{\rm optc}}}{\gamma_{\rm opta}+\gamma_{\rm optc}-i\Omega}\hat b_{\rm th}\approx i\sqrt{\frac{\gamma_m}{\gamma_{\rm opt}}}\hat b_{\rm th},\\
&\hat a_{\rm out}^{\rm th}(\Omega)=-\frac{2i\sqrt{\gamma_m\gamma_{\rm opta}}}{\gamma_{\rm opta}+\gamma_{\rm optc}-i\Omega}\hat b_{\rm th}\approx -i\sqrt{\frac{\gamma_m}{\gamma_{\rm opt}}}\hat b_{\rm th},
\end{split}
\ee
where the second equality comes from the setting $\gamma_{\rm opta}=\gamma_{\rm optc}=\gamma_{\rm opt}$.

For a small enough thermal noise that does not cause serious quantum decoherence, we need to satisfy:
\be
\frac{\gamma_m}{\gamma_{\rm opt}}\frac{k_B T}{\hbar \omega_m}\ll S_{a_{\rm in}a_{\rm in}}\Rightarrow \frac{T}{Q_m}\ll
\hbar \gamma_{\rm opt} S_{a_{\rm in}a_{\rm in}},
\ee
where $S_{a_{\rm in}a_{\rm in}}$ is the spectrum of quantum fluctuation of input squeezed states.

\begin{figure}[htbp]
   \centering
   \includegraphics[width=3.in]{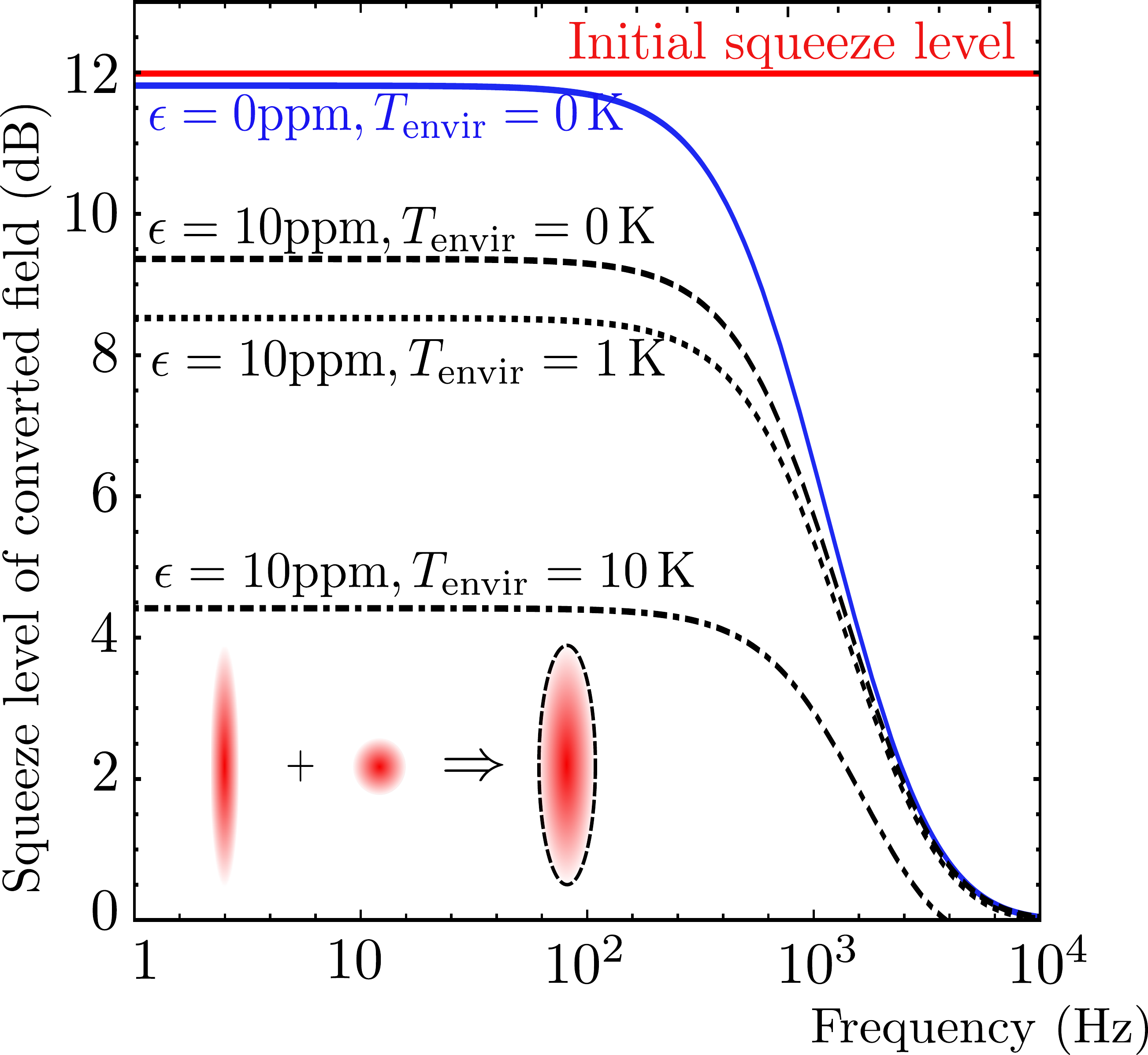} 
   \caption{Squeeze level for the converted field vs frequency. We set 12\,dB initial squeezing being filtered by the OMFC. The squeeze degradation at high frequency is mainly due to the cavity bandwidth, while thermal and optical loss dominates the degradation at low frequency.}
   \label{fig:omfc_sqz}
\end{figure}

Besides, the $\hat c_{\rm in}$ term in Eq.(8) ($\propto \Omega/\gamma_{\rm opt}$) and the ignored far off-resonant sideband fields ($\propto \gamma_{a,c}/\omega_m$) will also decohere the original quantum states. However, their effects are typically much smaller than the thermal noise, as long as we have a strong optical damping factor $\gamma_{\rm opt}$ and the system is in the resolved-sideband limit. The other important decoherence channel is the optical loss of the optomechanical filter cavity. In Fig.\,\ref{fig:omfc_sqz}, the effect of loss and thermal noise on the squeeze level of the converted fields is shown.

\section{GWD configurations with frequency converter}
Now in this section, we discuss how to implement the above OMFC to improve the sensitivity of gravitational wave detectors. In this work, two different configurations are discussed: (1) using OMFC to generate a FD squeezed light for broadband  quantum noise reduction; (2) using OMFC to generate a FD rotation of the dark port output light, so that the radiation pressure noise can be evaded by variational readout.

\begin{figure}[htbp]
   \centering
   \includegraphics[width=3.4in]{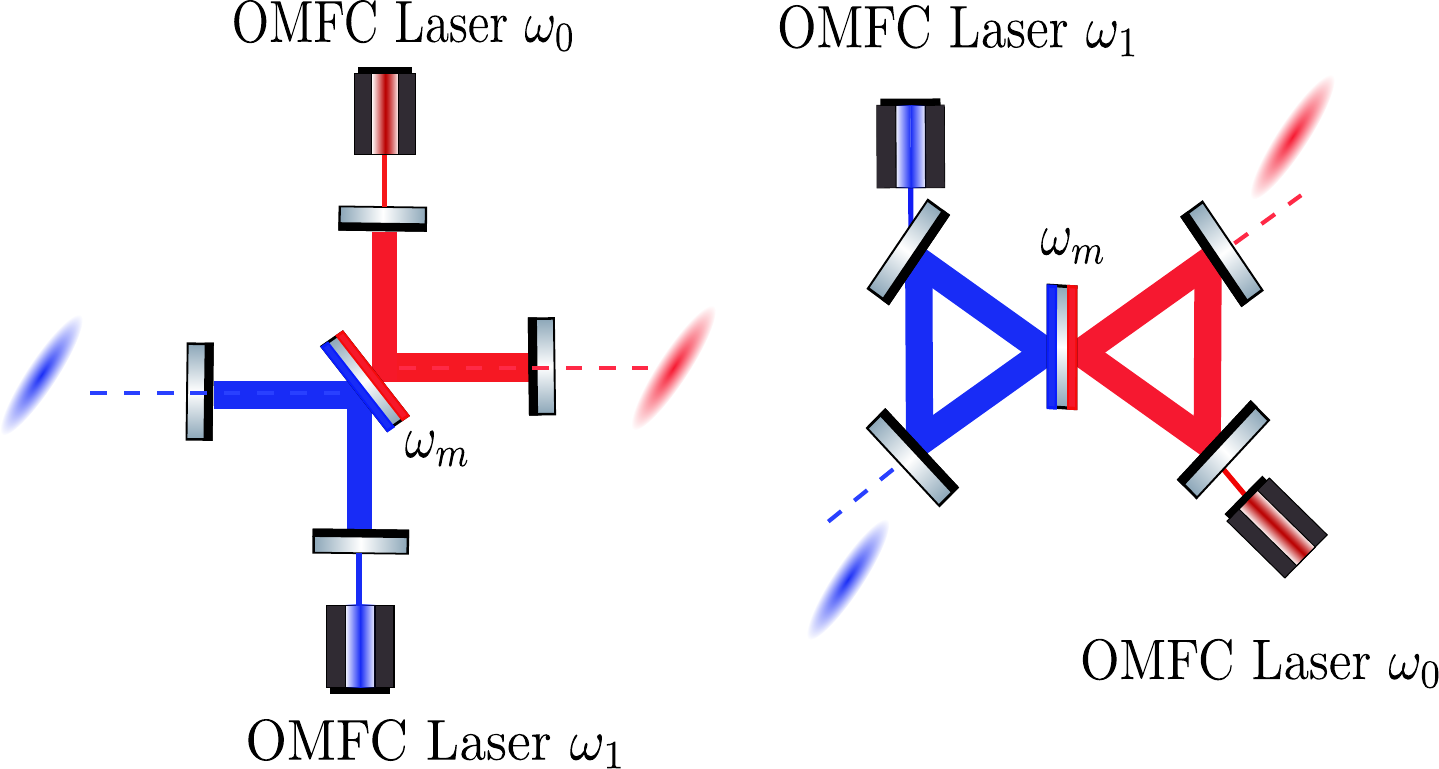} 
   \caption{Two example designs of OMFC where the pumping fields are separated from the probe fields at the injection port. Note that for the triangle cavity design, there is an additional bonus that the optical field propagating into the device from the output port will not participate in the optomechanical interaction, since the pumping field propagates in just the opposite way. Therefore it may help us to evade the back-scattering light when we imbed the OMFC into the GW detector design.}
   \label{fig:omfc_scheme}
\end{figure}

In Fig.\,\ref{fig:omfc_scheme} we show some design schemes for the OMFC that can be embedded into the gravitational wave detectors. The key of these designs is to separate the pumping field from the weak probe field being converted, which comes from the main interferometer, as we can see in later Fig.\,\ref{fig:fd_scheme} and Fig.\,\ref{fig:vr_scheme}. The final choice of the OMFC configurations depends on a more detailed experimental-based analysis. Table\,\ref{tab:parameters} gives some sample parameters for an optomechanical frequency converter.\\

\begin{table}[h]
   \centering  
    \begin{threeparttable}
\textbf{Optomechanical Frequency Converter}\vspace{5pt}
   \begin{tabular}{|ccc|}
   \hline 
 Symbols&Parameters & Values\\
  \hline
 $m$&oscillator mirror mass &1\,mg\\
$\omega_m/2\pi$&mechanical resonant frequency&1\,MHz\\
 $Q_m$ & mechanical quality factor & $5\times 10^7$\\
 $L_{a,c}$ & cavity length & $$ 1\,m\\
 $\gamma_{a,c}$ & cavity bandwidth & $1.5\times10^5 $\,rad/s\\
 $P_{a,c}$ & resonating power & 170\,W\\
 $T_{\rm envir}$& environmental temperature &1 K\\
 $\epsilon$ & cavity round-trip optical loss & 10 {\rm ppm}\\
 \hline
 \end{tabular} 
  \centering
\textbf{Main Laser Interferometer}\vspace{5pt}
  \begin{tabular}{|ccc|}
   \hline 
 Symbols&Parameters & Values\\
  \hline
 $M$ & test mirror mass &40\,kg\\
 $L_{\rm arm}$ & arm cavity length & 4\,km\\
 $T_{\rm ITM}$ & input test mass transmission & 0.014\\
 $P_{\rm arm}$ & arm resonating power & 800\,kW\\
 $\epsilon_{\rm circ}$ & single-trip circulator loss& 0.5\%\\
 $r$ &input squeezing level &12 \,dB\\
 $T_{\rm SRM}$ &signal recycling mirror transmission & 0.35\\
 $\epsilon_{\rm ext}$& external loss &0.5\%\\
 $\Delta$& Frequency difference& $\sim15$\,MHz\\
 \hline
 \end{tabular}
\caption{Sample Parameters for optomechanical frequency converter and the laser interferometer.}
   \label{tab:parameters}
   \end{threeparttable}
\end{table}

\subsection{Frequency dependent squeezing using OMFC.}
\begin{figure}[h]
   \centering
   \includegraphics[width=3.5in]{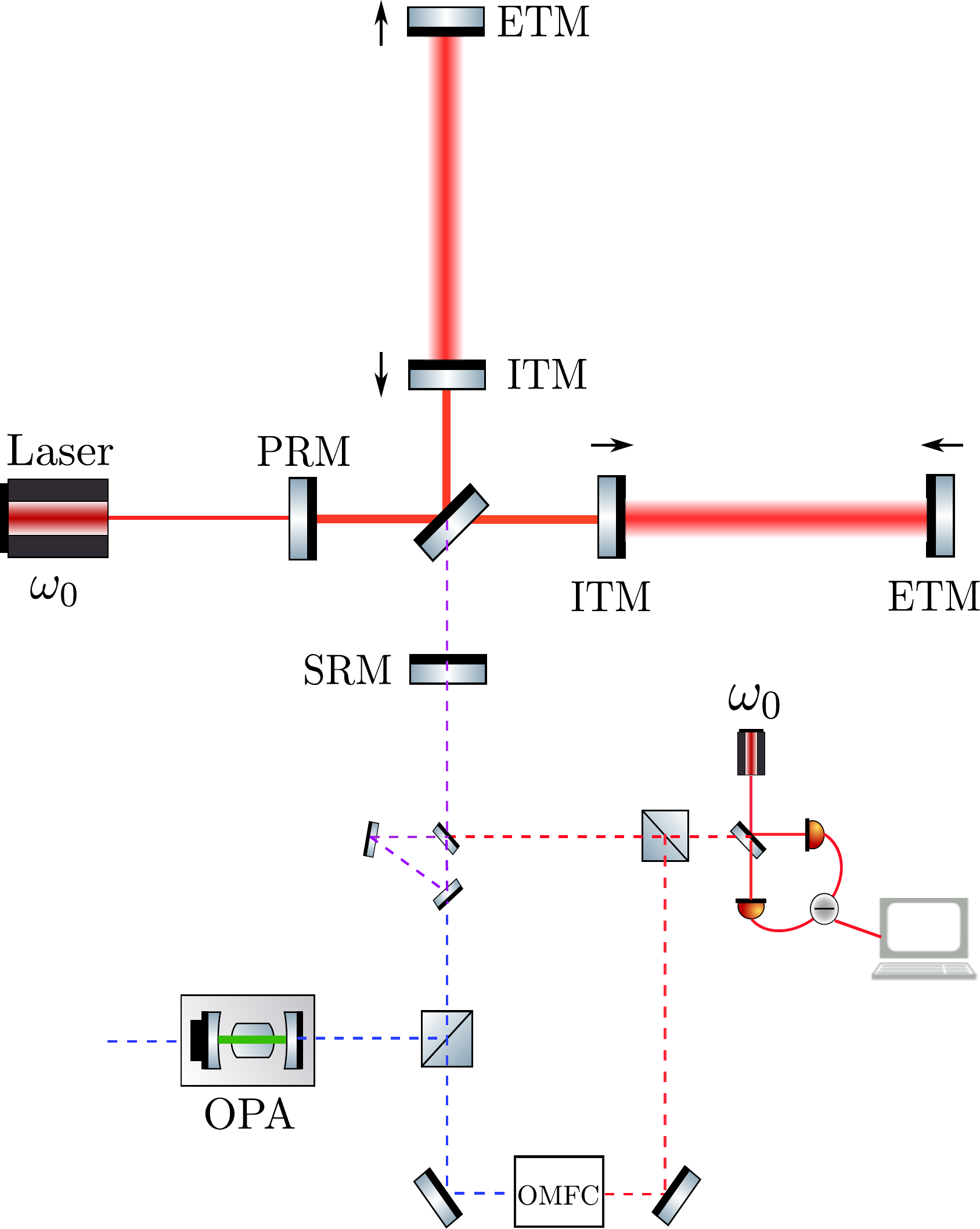} 
   \caption{Broadband squeezing assisted by arm cavity filtering and OMFC:  a conceptual design configuration.}
   \label{fig:fd_scheme}
\end{figure}

\begin{figure}[h]
   \centering
   \includegraphics[width=3.5in]{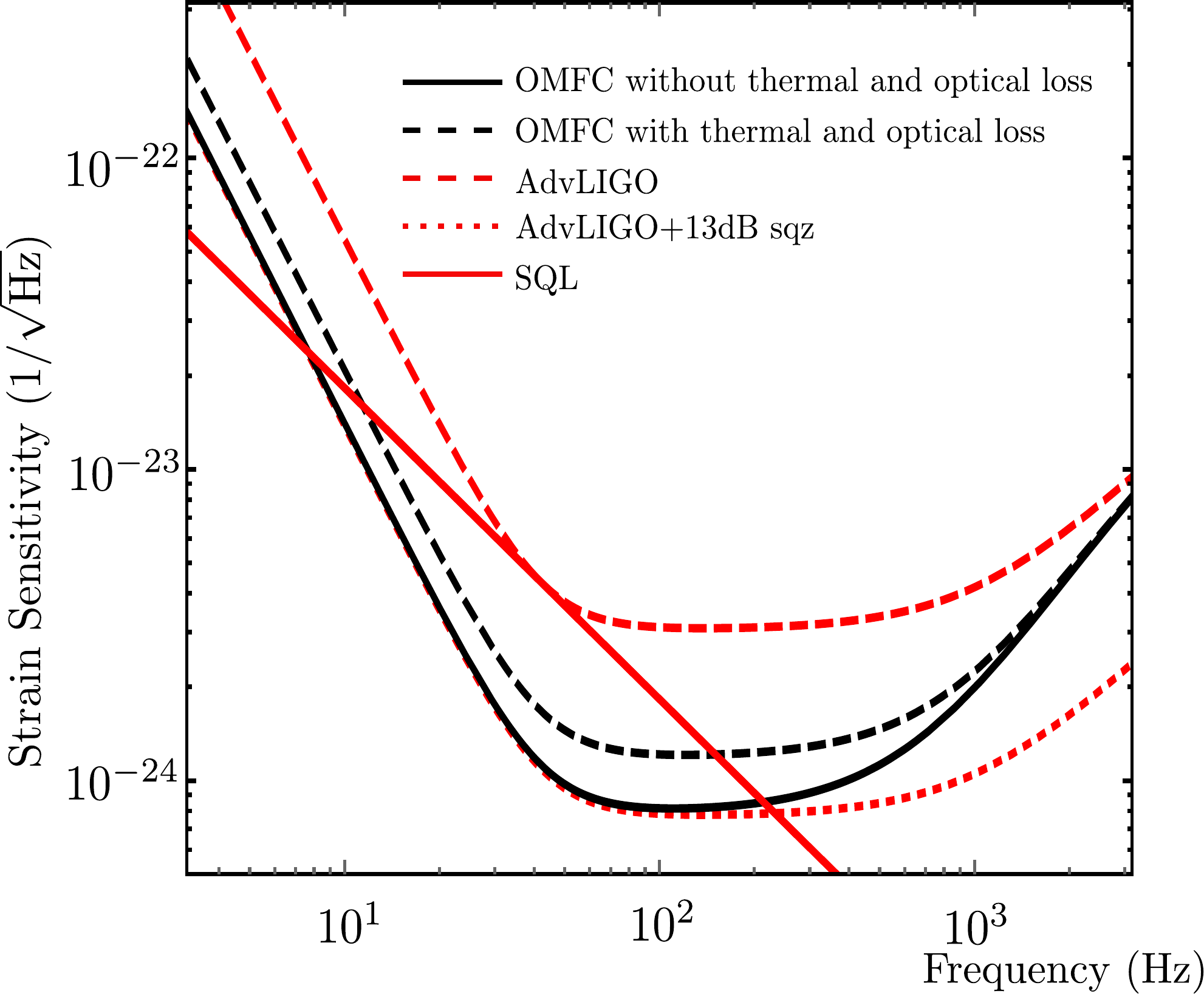} 
   \caption{Sample sensitivity of gravitational wave detector configuration shown in Fig.\,\ref{fig:fd_scheme}. Parameter choice follows Table\,\ref{tab:parameters}.}
   \label{fig:fd_sensitivity}
\end{figure}

The scheme of frequency dependent squeezing using OMFC is shown in Fig.\,\ref{fig:fd_scheme}. The working principle can be described as follows: (1) a squeezed light is injected into the main interferometer, with centre frequency $\omega_a$ far detuned away from the carrier frequency $\omega_0$ of the main interferometer. This field neither carries GW signal nor drives the test masses motion. Therefore the interferometer behaves as a simple empty cavity. Properly tuning the interferometer parameters can generate a required FD squeeze angle\,\cite{Ma2017}. (2) This frequency dependent squeezed light centred at $\omega_a$, then be injected into the OMFC, and be converted to ideally the same FD squeezed light centred at $\omega_c=\omega_0$. (3) This $\omega_c$-centred FD squeezed light is injected into the main interferometer again, carrying the gravitational wave signal and beating the broadband quantum noise. The sensitivity curve using the parameters in Table.\,\ref{tab:parameters} is shown in Fig.\,\ref{fig:fd_sensitivity} with sample parameters given in Table\,\ref{tab:parameters}. \\

When using OMFC to do FD squeezing, the criterion Eq.\,\eqref{eq:criterion} will be modified as:
\be
\frac{T}{Q_m}\ll \frac{\hbar\gamma_{\rm opt}e^{-2q}}{k_B},
\ee
where $S_{a_{\rm in}a_{\rm in}}=e^{-2q}$ is the noise spectrum of the squeezed quadrature. For a 13 dB squeezed light, we have $e^{-2q}\approx 0.05$, leads to $T/Q_m\ll 5\times10^{-13}\gamma_{\rm opt}$. Using parameters in the Table\,\ref{tab:parameters}, the $\gamma_{\rm opt}\approx 10^5$ and $T/Q_m\ll 5\times 10^{-7}$\,K. Note  there is in principle no problem to increase the value of $\gamma_{\rm opt}$, unlike the case in\,\cite{Ma2014} where $\gamma_{\rm opt}$ is constrained by the frequency that the sensitivity touches the SQL.

The implementation of the scheme could use the frequency difference $\omega_a-\omega_c$ $\sim100$ MHz. In this case, the interferometer optics can be kept almost the same because the bandwidth of reflective coating is much wider than 100MHz, and the polarisation directions of optical fields inside the interferometer are the same.

There could be back-scattering light from interferometer coming into the OMFC cavity. However, If we use the triangular cavity OMFC as in Fig.\,\ref{fig:omfc_scheme}, the back-scattering light coming into the OMFC will propagate in the opposite direction as the OMFC pumping light and have minimum opto-mechanical interactions converts back as noise contamination.  Therefore the scheme here can in principle evade the back-scattering noise.

\subsection{Output filtering using OMFC.}
\begin{figure}[htbp]\label{fig:scheme}
   \centering
   \includegraphics[width=3.5in]{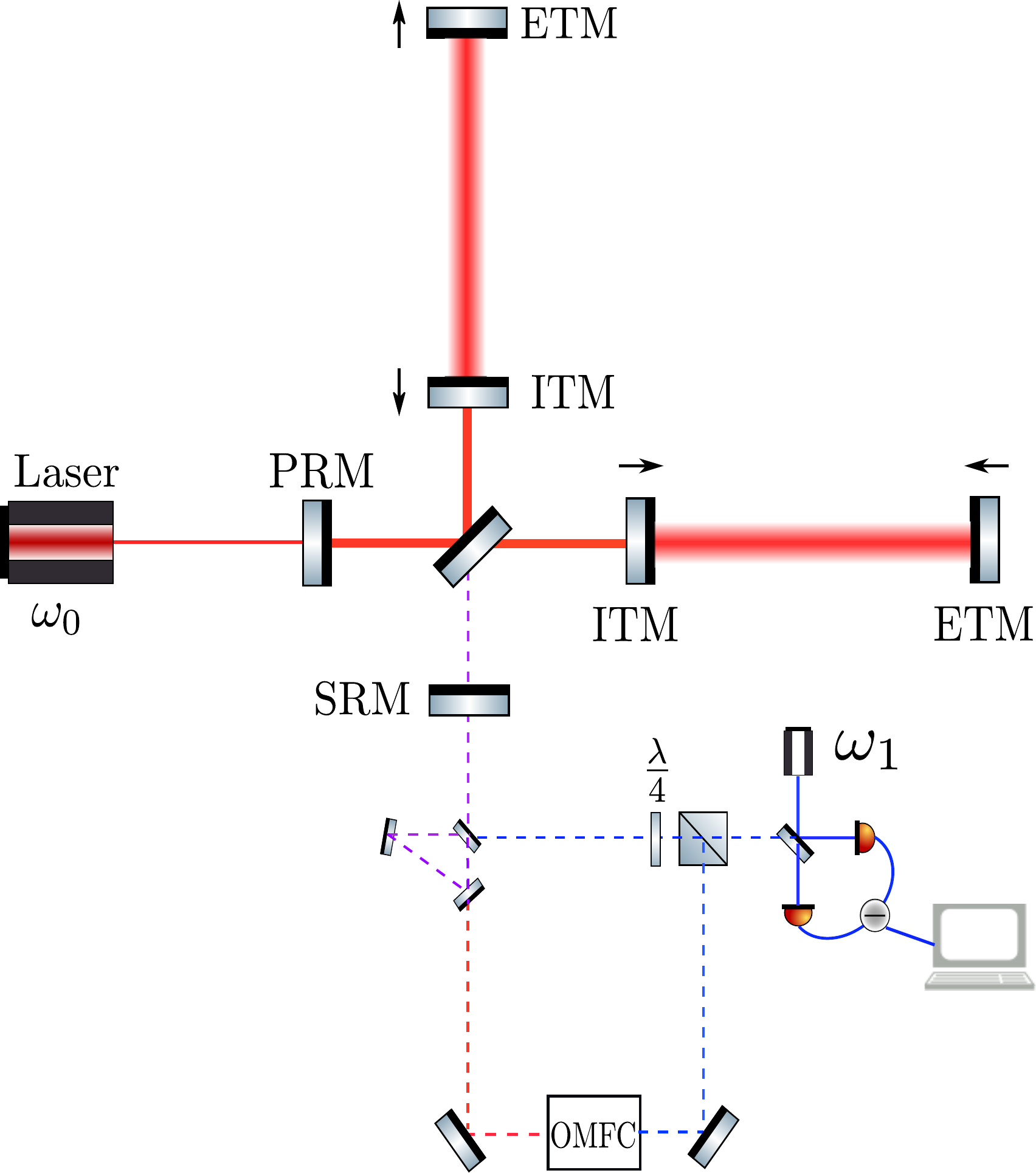} 
   \caption{Back-action evasion by OMFC assisted variational readout scheme: a conceptual design configuration.}
   \label{fig:vr_scheme}
\end{figure}

The scheme of variational readout using OMFC is shown in Fig.\,\ref{fig:vr_scheme}, the working principle is relatively simpler compared to FD squeezing. Basically, the output field from the interferometer centred around $\omega_0$ will be directly converted to field around another frequency which is far detuned from $\omega_0$. This field will then be injected back into the interferometer and sees a frequency dependent rotation.

When using OMFC to do variational readout, the criterion Eq.\,\eqref{eq:criterion} will be:
\be
\frac{T}{Q_m}\ll \frac{\hbar\gamma_{\rm opt}}{k_B}.
\ee
As what we proposed in the FD squeezing scheme, if the  parameters takes the value of Table\,\ref{tab:parameters}, the $\gamma_{\rm opt}\approx 10^5$. Finally we have $T/Q_m\ll 5\times 10^{-6}$\,K, which is relatively easier to be achieved. One can also further increase the intracavity power $P_{a,c}$ to relieve this condition.

\begin{figure}[htbp]\label{fig:scheme}
   \centering
   \includegraphics[width=3.3in]{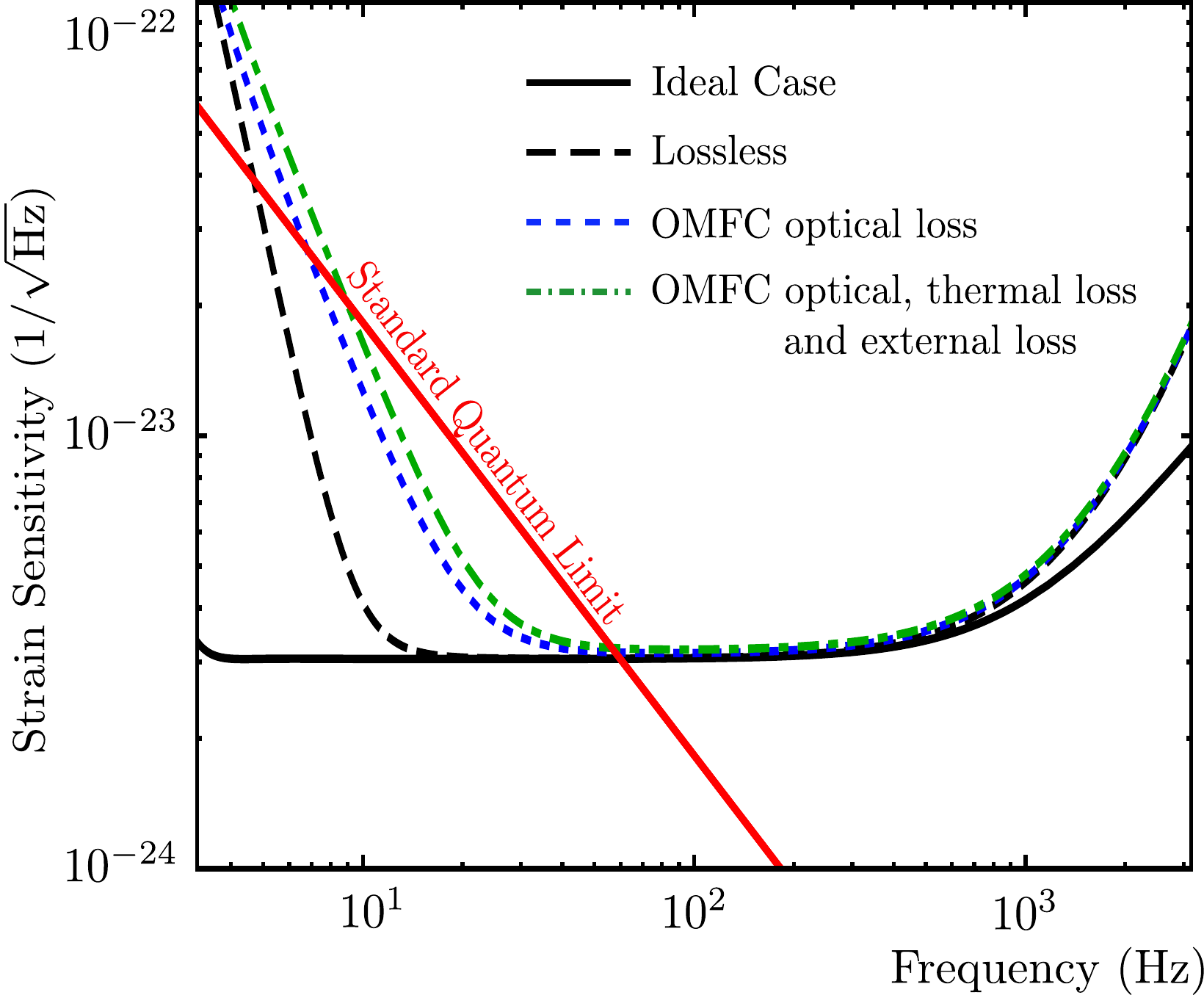} 
   \caption{Sample sensitivity of gravitational wave detector configuration shown in Fig.\,\ref{fig:vr_scheme}. Parameter choice follows Table\,\ref{tab:parameters}}
   \label{fig:vr_sen}
\end{figure}

The variational readout scheme is similar to the FD squeezing scheme. The signal light from the interferometer enters OMFC after passing through the output mode cleaner. The converted light from the OMFC maintains the signal information. Its polarisation direction is rotated 45 degrees before injected into the interferometer through the Faraday rotator, so that inside the interferometer it will have the same polarisation as that of the carrier light. The sensitivity curve is given in Fig.\,\ref{fig:vr_sen}.

\subsection{Effect of imperfections of OMFC to the sensitivity}
The sensitivity degradations shown in Fig.\,\ref{fig:fd_sensitivity} and Fig.\,\ref{fig:vr_sen} contributed by the imperfections of OMFC system mainly contain three pieces:  above all, the angle error created by OMFC cavity,  then the loss of the OMFC cavity and the thermal noise. In particular, for variational readout scheme using OMFC, the angle error contributed degradation is very serve because of the strong back-action noise at low frequencies.

The almost perfect conversion formula Eq.\eqref{eq:conversion} is only an approximate formula. In reality, the OMFC not only contributes conversion, but also induces a tiny rotation to the field being converted, and this rotation needs to be accounted in choosing the parameters for the interferometer as a filter cavity. The angle error associated with this additional rotation is the most severe degradation.

\begin{figure}[htbp]
   \centering
   \includegraphics[width=3in]{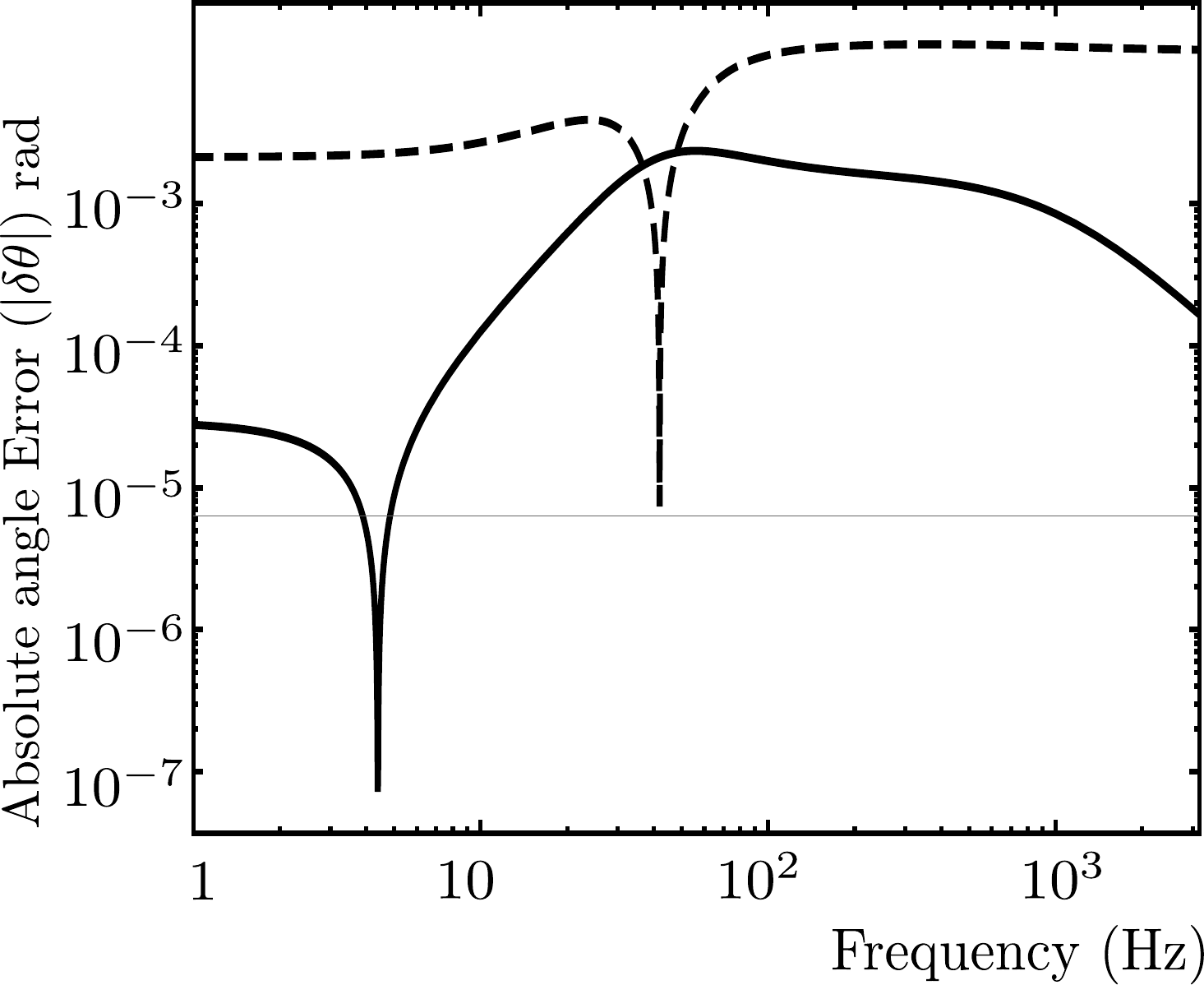} 
   \caption{The homodyne angle error vs frequency. The black solid curve is the angle error of an ideal case when there is no OMFC imperfections while the black dashed curve is the angle error of the real case after we optimise the filter parameters. The very sharp peaks here are due to the fact that we are plotting the absolute value of angle error. }
   \label{fig:angle_error}
\end{figure}

The exact formula for the conversion rate derived from the Hamiltonian is given as:
\be
\frac{\hat c_{\rm out}}{\hat a_{\rm in}}=\frac{\gamma_{\rm opt}(1+\epsilon_2+i\epsilon_1)/(1-i\epsilon_3)^2}{-i\Omega(1+\epsilon_2+i\epsilon_1)+\gamma_{\rm opt}/(1-i\epsilon_3)},
\ee
where $\epsilon_1=\gamma/2\omega_m$, $\epsilon_2=\Omega/2\omega_m$, $\epsilon_3=\Omega/\gamma$. To the leading order, it can be approximated as:
\be
\begin{split}
\frac{\hat c_{\rm out}}{\hat a_{\rm in}}\approx\left(\frac{\gamma_{\rm opt}}{\gamma_{\rm opt}-i\Omega}\right)&\left[1+\frac{\gamma_{\rm opt}}{\gamma_{\rm opt}-i\Omega}(\epsilon_2+i\epsilon_1)\right.\\
&\left.\qquad+\frac{i\gamma_{\rm opt}+2\Omega}{\gamma_{\rm opt}-i\Omega}\epsilon_3\right].
\end{split}
\ee
At very low frequency, the correction to the perfect conversion is:
\be
\frac{\hat c_{\rm out}}{\hat a_{\rm in}}\approx1+i\epsilon_1\approx \cos{\epsilon_1}+i\sin{\epsilon_1},
\ee
where $\epsilon_1=\gamma/(2\omega_m)\approx\sin\delta\theta$ as a small rotation angle $\delta \theta$. Using the parameters in Table\,\ref{tab:parameters}, we estimate the $\epsilon_1\sim 10^{-2}$\,rad and corresponding sensitivity degradation at $3$\,Hz is around $2.6$dB. However, angle error of $\sim 10^{-2}$ can be decreased by optimising other filter cavity parameters (see the Appendix) and we find that the optimal sensitivity degradation can be in principle reduced to $1.7$dB. In Fig.\,\ref{fig:angle_error}, we show the residue angle error after we optimise the filter parameters, compare to the ideal case.

As shown in Fig.\,\ref{fig:vr_sen}, the effect of optical loss at low frequency region is significant. 
This optical loss rate can be effectively estimated as: 
\be
\epsilon_{\rm OMFC}\approx\frac{c\epsilon}{L_{a,c}\gamma_{a,c}}\frac{\gamma_{\rm opt}^2}{\gamma_{\rm opt}^2+\Omega^2}.
\ee
Substituting the parameters in Table\,\ref{tab:parameters}, we have $\epsilon_{\rm OMFC}\approx 0.05$. Plug this loss rate into Eq.\eqref{eq:loss_sen}, resulted a $\sim 10$ times degradation to the shot noise level at $10$\,Hz--- a very good estimation to the result in Fig.\ref{fig:vr_sen}.

The contribution of thermal noise at low frequency region can be estimated as:
\be
S_{\rm th}\approx\frac{8k_BT_{\rm envir}}{\hbar\gamma_{\rm opt}Q_m}\frac{\gamma_{\rm opt}^2}{\gamma_{\rm opt}^2+\Omega^2},
\ee
and with our parameters, its value is roughly $\sim 0.2$, its correction is only slightly above the shot-noise level.

\section{Discussion and Conclusion.}
In this work, combining the idea of a dual-use laser interferometer, we discussed the application of optomechanical frequency converter in improving the sensitivity of gravitational wave detectors, specifically, in achieving the frequency dependent squeezing and variational readout. The effect of imperfections of optomechanical devices to the final sensitivity is carefully analysed. We found that comparing to the previous proposals of using the peculiar dispersion of optomechanical device to do squeeze filtering\,\cite{Ma2014} and bandwidth enhancement\,\cite{Miao2015PRL}, the damping rate of the optomechanical frequency converter does not have the limitations so that it can be made large to dilute the effect of thermal noise. 
From the technical aspect, the optomechanical device based on a triangle cavity design can evade the back-scattering noise. We also pointed out that the imperfections of optomechanical devices have a rather significant degradation to the low frequency sensitivities in the variational readout scheme.

It is worth to mention that the application of frequency conversion concept in gravitational wave detectors is not limited by the platform of optomechanical device. In principle, a crystal-based frequency conversion can also be made. Notably, experimental demonstration of frequency conversion of squeezed light using crystals has been realised in\,\cite{Vollmer2014}. Comparing to the crystal based frequency converter, optomechanical frequency converter has its own advantages and also disadvantages. 

Full optical based crystal frequency converter (such as\,\cite{Vollmer2014}) is suitable for conversion between the light with two very different frequencies. However, the optical frequency window of the interferometer is limited. Therefore, this design may have significant interferometer loss issue. Ideas of using acoustic-optic modulator (AOM) to convert squeezed light between MHz separated frequencies are recently raised\,\cite{Roman_priviate} and needs more in-depth analysis, which could be useful for future application. However, such AOM based schemes may still have the problem of back-scattering noise since the acoustic driving of the crystals does not distinguish different propagation directions.  Moreover, optomechanical devices can provide us with advantages of tunability.

In this paper, by combining the idea of optomechanical frequency conversion and arm cavity filtering, we discussed the implement of frequency converter to beat the standard quantum limit of gravitational wave detectors. The feasibility, though is not strongly constrained by the value of optomechanical cooling rate $\gamma_{\rm opt}$ as in\,\cite{Ma2014,Miao2015PRL}, still depends on the future technology of low loss optics and mechanics. 

\section*{acknowledgement}
The authors thank M. Korobko, F. Ya. Khalili and Stefan Danilishin for useful discussions and comments. Y. M. also thank R. Schnabel for commenting the crystal frequency converter. Y. C. and Y. M are supported by the National Science Foundation through Grants PHY-1612816, PHY-1708212, and PHY-1708213, the Brinson Foundation, and the Simons Foundation (Award Number 568762). C. Z. is supported through Australian Research Council (ARC) Centre of Excellence for  Gravitational Wave Discovery Project CE170100004.

\appendix
\section{Effect of rotation error and loss on variational readout}
The principle of variational readout is as follows. Suppose the in-out relation of an interferometer can be written as:
\be
\begin{split}
&\hat b_1=e^{2i\beta}\hat a_1,\\
&\hat b_2=e^{2i\beta}(\hat a_2-\kappa\hat a_1)+e^{i\beta}\sqrt{2\kappa}\frac{h}{h_{\rm SQL}},
\end{split}
\ee
where $\kappa=16\omega_0 I_c \gamma_{\rm ifo}/(M\Omega^2(\Omega^2+\gamma_{\rm ifo}^2)L_{\rm arm}c)$ describe the pondermotive effect in the main interferometer. The $I_c, \gamma_{\rm ifo}, M, L_{\rm arm}$ are the main interferometer intra-cavity power, mass of test mass mirror and arm length, respectively. $\beta$ is the phase for the field accumulated inside the interferometer. The $h_{\rm SQL}=8\hbar/m\Omega^2L_{\rm arm}^2$ is the standard quantum limit.

Combine these two quadratures in a frequency-dependent way, we obtain:
\be
\begin{split}
\hat b_\theta&=\hat b_1\sin\theta_\Omega+\hat b_2\cos\theta_\Omega,\\
&=e^{2i\beta}(\sin\theta_\Omega-\kappa\cos\theta_\Omega)\hat a_1+e^{2i\beta}\cos\theta_\Omega\hat a_2+{\rm signal}.
\end{split}
\ee
Properly chosen the homodyne angle so that $\tan\theta_{\rm vr}=\kappa$, we can completely evade the radiation pressure noise. However, if there is some angle error, the remnant radiation pressure noise term is given as:
\be
\begin{split}
\delta \hat b_\theta&\approx e^{2i\beta}(\cos\theta_{\rm vr}+\kappa\sin\theta_{\rm vr})\delta\theta\hat a_1\\
&=e^{2i\beta}(1+\kappa^2)\cos\theta_{\rm vr}\delta\theta\hat a_1.
\end{split}
\ee
Since at low frequency region, the $\kappa$ is very large, therefore even a very small angle error will create a significant effect (for example, at $3.1$\,Hz$, \kappa^2\sim4.5\times10^4, \delta\theta\sim 10^{-5}$\,rad and $S_{\delta \hat b\delta \hat b}\sim 2\times 10^{-5}/{\rm Hz}$, while at the same time, the shot noise level at 3.1\,Hz $S_{\rm shot}=\cos^2\theta_{\rm vr}\sim 10^{-7}/{\rm Hz}$). Using the parameters of main interferometer of our scheme, the angle error at low frequency region and its effect on the sensitivity curve is shown in Fig.\,\ref{fig:angle_error}.

For the effect of loss, let us effectively describe the in-out relation contain loss noise as:
\be
\begin{split}
&\hat b_1=\sqrt{1-\epsilon}e^{2i\beta}\hat a_1+\sqrt{\epsilon}\hat n_1,\\
&\hat b_2=\sqrt{1-\epsilon}\left[e^{2i\beta}(\hat a_2-\kappa\hat a_1)+e^{i\beta}\sqrt{2\kappa}\frac{h}{h_{\rm SQL}}\right]+\sqrt{\epsilon}\hat n_2.
\end{split}
\ee
Combining them using variational readout scheme, we have:
\be
\begin{split}
\hat b_\theta=&\sqrt{1-\epsilon}e^{2i\beta}\hat a_2\cos\theta_{\rm vr}+\sqrt{\epsilon}(\hat n_1\sin\theta_{\rm vr}+\hat n_2\cos\theta_{\rm vr})\\
&+\sqrt{1-\epsilon}e^{i\beta}\sqrt{2\kappa}\frac{h}{h_{\rm SQL}},
\end{split}
\ee
and the sensitivity is given by:
\be\label{eq:loss_sen}
S_{hh}=\frac{h_{\rm SQL}^2}{2\kappa}\left[1+\frac{\epsilon}{(1-\epsilon)\cos^2\theta_{\rm vr}}\right].
\ee
As an estimation, for $\epsilon\sim0.5\%$ at frequency $\sim 3.1$\,Hz, we have $\cos^2\theta_{\rm vr}\sim 3\times 10^{-4}$, therefore the loss effect is roughly $160$ times larger than the shot noise level. This simple estimation shows that loss effect strongly degrades the efficiency of variational readout scheme. Moreover, in the OMFC assisted variational readout scheme, the external loss must also include the loss of the OMFC and thermal noise.

\section{Tuning of frequency dependent angle}
For a simple cavity which rotate a squeezed light with centre frequency detuned from cavity resonant by $\Delta$, the rotation angle $\xi$ is given by\,\cite{Miao2014}:
\be
\begin{split}
\tan\xi(\Omega)=\frac{2\Omega\gamma}{\Delta^2-\Omega^2+\gamma^2}
\end{split}
\ee
If there is a angle error $\delta\xi$, the correction to the leading order is given as:
\be
\tan\xi'(\Omega)\approx\tan\xi(\Omega)\left[1-\tan\xi(\Omega)\left(\frac{\delta\Delta}{\Omega\gamma}\right)\right]
\ee
Since we have the simple trig identity:
\be
\tan(\xi+\delta \xi)=\frac{\tan\xi+\tan\delta\xi}{1-\tan\xi\tan\delta\xi}\approx\tan\xi(1+\delta\xi\tan\xi),
\ee
therefore we have:
\be
\delta\xi\approx\frac{\delta\Delta}{\Omega\gamma}.
\ee
as an estimation, for compensate a angle error $\sim 10$\,mrad, with $\Delta\sim\gamma$ and at low frequency region (where the radiation pressure effect is mostly strong, and let us just take $\Omega\sim 2\pi\times 1$\,rads/s as an example),  the detuning compensation would be: $\delta\sim 10$\,mrad/s. Another probably easier method to solve this problem is to keep the original detuning but set a DC homodyne angle offset to compensate this angle error.

\bibliographystyle{unsrt}
\bibliography{yiqiubib}

\end{document}